\documentclass{article}
\usepackage{hiph-art}
 \volnumber{00} \issuenumber{0} \edyear{2005}                             
 \frompage{000} \topage{000}                                              
\recrevdate{\today}                                              
\def\rightmark{Thermodynamics
of the quark-gluon plasma}
\def\leftmark{A. Rebhan}

\def\0{\over } 
\def\2{{\textstyle{1\over2}}} \def\4{{\textstyle{1\over4}}}
\def\5{\hat } \def\6{\partial }
\def\8#1{{\textstyle{#1}}}

\newcommand{\bea}{\begin{eqnarray}}
\newcommand{\eea}{\end{eqnarray}}
\newcommand{\be}{\begin{equation}}
\newcommand{\ee}{\end{equation}}
\newcommand{\nn}{\nonumber\\ }
\newcommand{\muMS}{\bar\mu_{{\rm MS}}}
\newcommand{\g}{g_{\rm eff}}

\title{Weak-coupling techniques for\\[2pt] 
the thermodynamics of the quark-gluon plasma} 
\authors{
{Anton Rebhan
\index{Rebhan, A.} 
}\\[2.812mm]
{\normalsize
Institute for Theoretical Physics, Technical University Vienna\\ 
Wiedner Hauptstr. 8-10, A-1040 Vienna, Austria}\\[0.2ex] 
}
 
\abstract{We describe some of the recent progress in the calculation
of thermodynamic quantities in QCD at high temperatures and densities
by weak-coupling techniques and extrapolation to realistic coupling strength.
We argue that a (mostly) weakly coupled quark-gluon plasma 
at temperatures only a few times the transition temperature
is not yet ruled out by the observed fast apparent thermalization
at RHIC, as nonabelian plasma instabilities might provide an efficient
mechanism for fast isotropization even in a collisionless plasma.}
 
\makeindex
\begin{document}
 
\maketitle

\section{Introduction}\label{intro}

To approach
the thermodynamics of the strong interactions by weak-coupling techniques
\cite{Kap:FTFT,Blaizot:2003tw,Kraemmer:2003gd}
might appear to be a hopeless enterprise, at best of academic interest or
applicable only to 
such high temperatures and densities that it is completely irrelevant
for the hunt of the quark-gluon plasma in heavy-ion physics. 
However, while the physics of the deconfinement transition certainly cannot
be captured by Feynman diagrams of quarks and gluons, it may actually be
the case that 
extrapolations from asymptotically high temperatures already begin to make sense
quantitatively at temperatures only a few times the deconfinement temperature, 
and I shall present some evidence for this in what follows.
It is true that strict perturbation theory fails spectacularly, even at
temperatures as high as $10^5$ times the deconfinement temperature, where
perturbative QCD should work without difficulty. But this failure
has little to do with nonperturbative properties of 
nonabelian gauge theories ---
it even occurs in such trivial theories as scalar field theory in
the large-$N$ limit, 
where thermal quasiparticles are free of interactions.
Indeed, this failure can be repaired to some extent
(at least for bulk thermodynamics), as I shall describe below,
and the results from comparison with lattice results are quite encouraging.

\begin{figure}[htb]
\begin{center}\leavevmode
\epsfysize=4.1cm
\epsfbox{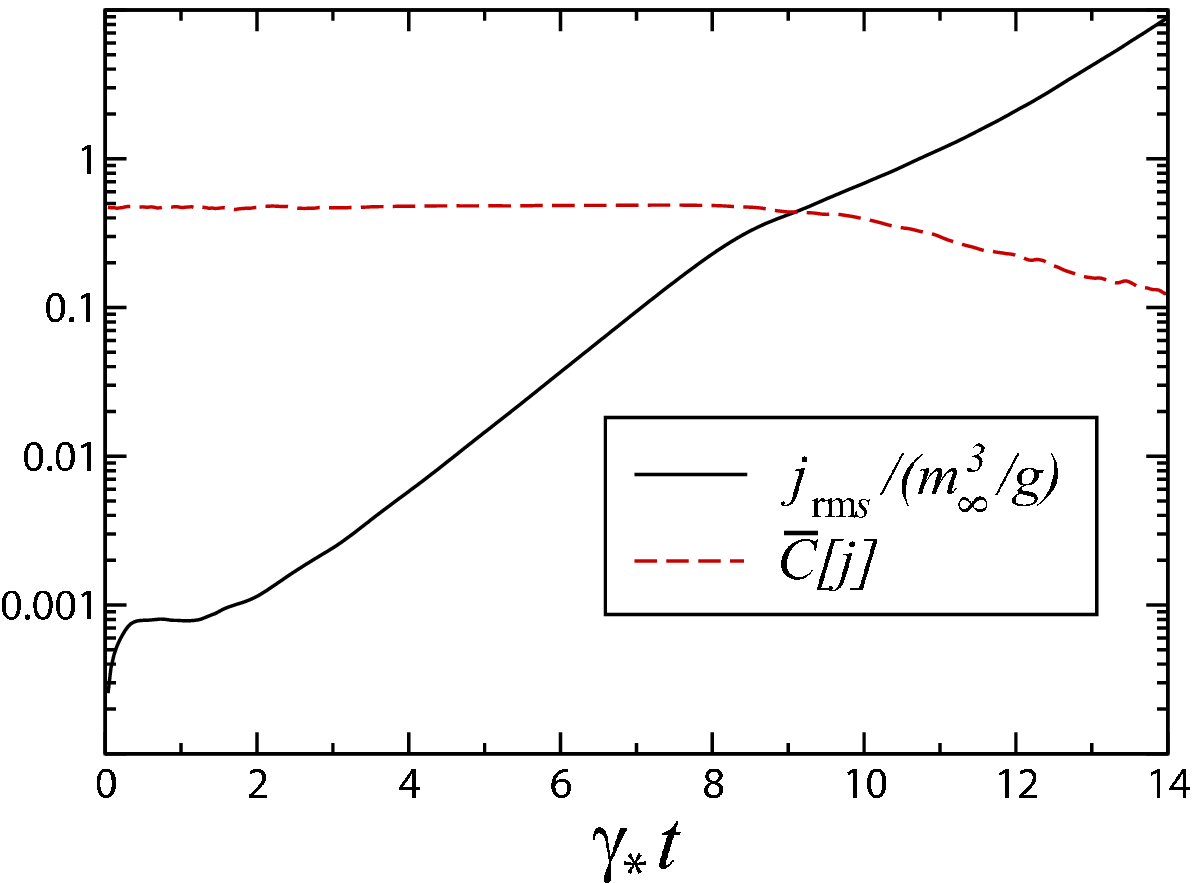}
\epsfysize=4.1cm
\epsfbox{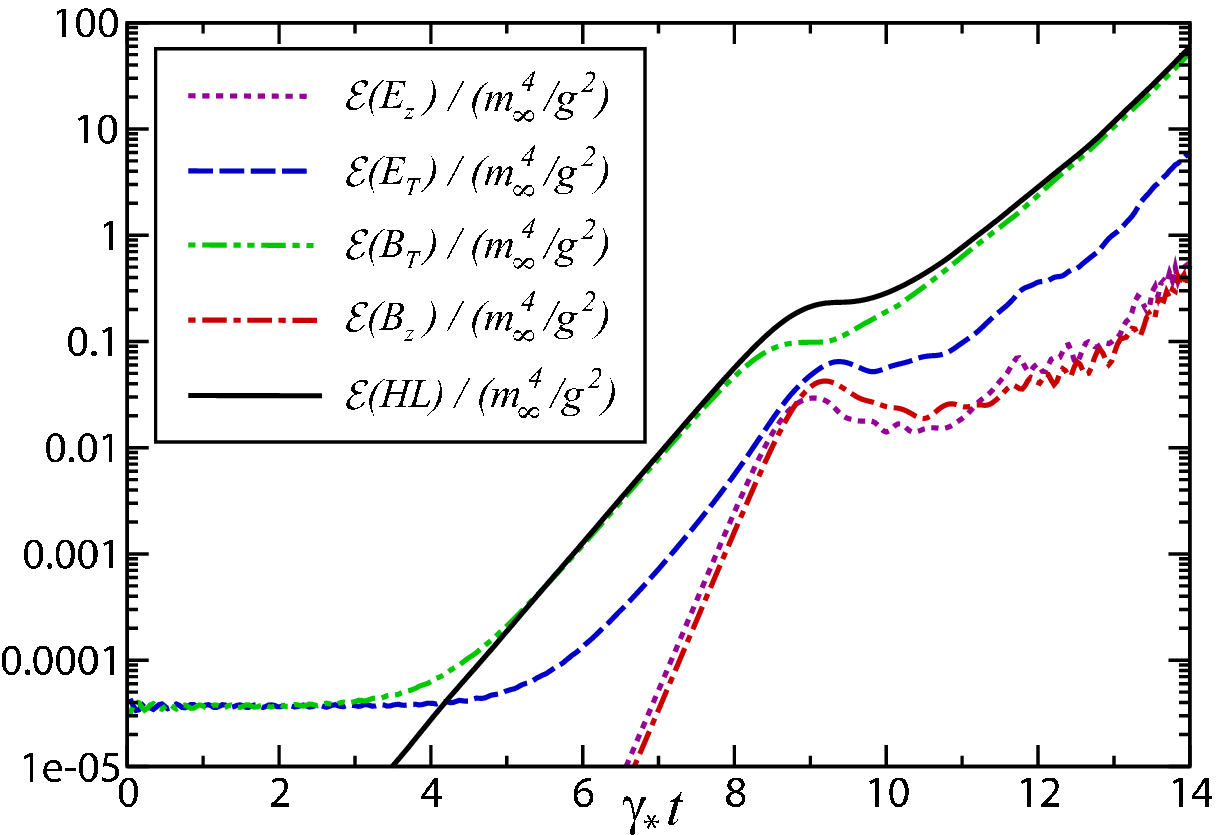}
\end{center}
\vspace*{-0.5cm}
\centerline{\footnotesize\it (a) \hspace{5.3cm} (b)}
\vspace*{-0.2cm}
\caption[]{{\it (a)} The exponential growth of nonabelian plasma instabilities
as measured by the rms value of the induced current density $j$ (full line)
and a measure of local abelianization ($\bar C$)
in the numerical simulation of Ref.~\cite{Rebhan:2004ur};
{\it (b)} the total energy transferred from hard modes into
the soft unstable modes (full line) and the energies in
the various transverse (longitudinal) chromoelectric (-magnetic)
field components.}
\label{figprl}
\end{figure}

It may be argued that the partial 
successes of weak-coupling techniques at temperatures
a few times the deconfinement temperature do not prove anything, and
that the quark-gluon plasma at such temperatures is essentially
strongly coupled. This is by now certainly the opinion of the majority
since the strongly coupled quark-gluon plasma has been heralded as
one of the major discoveries at RHIC. The evidence quoted for this
is the fast apparent thermalization and quick onset of hydrodynamic
behaviour, which seems incompatible with existing perturbative predictions
\cite{Wong:1996va,Baier:2000sb,Bjoraker:2000cf,Molnar:2001ux,Heinz:2004pj,Shuryak:2004kh,Wong:2004ik,Kovchegov:2005ss}.
However, it is not excluded that weak-coupling physics could
explain the fast apparent thermalization. Arnold et al.\ \cite{Arnold:2003rq,Arnold:2004ih,Arnold:2004ti}
have recently
pointed out that there is only evidence for fast isotropization, and
the latter may occur also in a weakly coupled plasma through
nonabelian plasma instabilities
as advocated since long by Mr{\'o}wczy{\'n}ski
\cite{Mrowczynski:1993qm,Mrowczynski:1994xv,Romatschke:2003ms}. 
Most recently, Romatschke, Strickland,
and myself \cite{Rebhan:2004ur} 
have produced supportive evidence for this picture by
numerical simulations of nonabelian plasma instabilities in the
nonlinear hard-loop approximation \cite{Mrowczynski:2004kv}. 
The latter is a weak-coupling framework,
but it nevertheless produces nonperturbative effects in the form
of nonperturbatively large fields which can efficiently wipe out
anisotropies in a collisionless plasma \cite{Dumitru:2005gp}. 
Fig.~\ref{figprl} shows
the exponential growth of 1+1-dimensional nonabelian plasma instabilities
(constant modes in the transverse plane) which continues
into the nonlinear regime so that it can only be stopped when
they begin to affect the trajectories of hard particles.

\section{Resumming perturbative results at high temperature}

\begin{figure}[htb]
\begin{center}\leavevmode
\epsfysize=4cm
\epsfbox{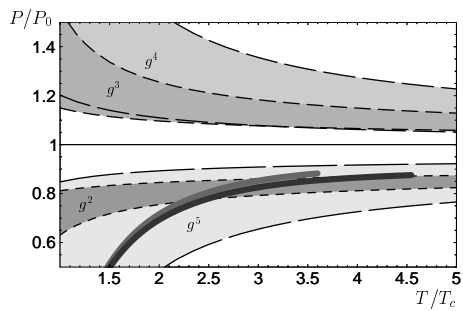}
\end{center}
\vspace*{-0.7cm}
\caption[]{Strictly perturbative results for the thermal pressure
of pure-glue QCD as a function
of $T/T_c$ (assuming $T_c/\Lambda_{\overline{\hbox{\scriptsize MS}}}=1.14$).
The various gray bands bounded by differently
dashed lines show the perturbative results
to order $g^2$, $g^3$, $g^4$, and $g^5$, 
using a 2-loop running coupling with $\overline{\hbox{MS}}$
renormalization
point $\muMS$ varied between $\pi T$ and $4\pi T$. The thick
dark-grey line shows the continuum-extrapolated
lattice results from reference~\protect\cite{Boyd:1996bx};
the lighter one behind that of a lattice calculation
using an RG-improved action \cite{Okamoto:1999hi}.}
\label{figqcd}
\end{figure}

However, even if this picture is not complete, and if there are
important nonperturbative effects, it is certainly worth knowing
how much of the physics of the quark-gluon plasma can already
be described by weak-coupling techniques.
The answer will certainly depend on the quantity under consideration,
and the easiest quantities should in fact be those related to the
equation of state, as the thermodynamic potential should be dominated
by contributions from hard modes, for which the effective coupling
should be smallest. Indeed, comparing perturbative results with
lattice results, the leading interaction contributions to the thermodynamic
potential $\propto g^2$ are doing quite well in reproducing the measured
deviation from ideal-gas behaviour for $T>T_c$, see Fig.\ref{figqcd}.

Problems arise from the next terms in the perturbative expansion,
where collective phenomena such as Debye screening come into play
\cite{Kap:FTFT}.
Up to and including order $g^5$ the thermal pressure of a quark-gluon
plasma with $N_f$ massless quarks reads
reads \cite{Arnold:1995eb,Zhai:1995ac,Braaten:1996jr}
\bea\label{Fpt}
 &&P = \8{8\pi^2\045}T^4 \biggl\{
\left(1+\8{21\032}N_f\right)-
\8{15\04}\left(1+\8{5\012}N_f\right){\alpha_s\0\pi}
+30\left[\left(1+\8{1\06} N_f\right)\left({\alpha_s\0\pi}\right)\right]^{3/2}\nn
&&\qquad\quad+\Bigl\{237.2+15.97N_f-0.413N_f^2+
\8{135\02}\left(1+\8{1\06} N_f\right)\ln\left[{\alpha_s\0\pi}(1+\8{1\06} N_f)\right]\nn
&&\qquad\qquad
-\8{165\08}\left(1+\8{5\012}N_f\right)\left(1-\8{2\033}N_f\right)
\ln{\muMS\02\pi T}
\Bigr\}\left({\alpha_s\0\pi}\right)^2\nn
&&\qquad\quad+\left(1+\8{1\06} N_f\right)^{1/2}\biggl[-799.2-21.96 N_f - 1.926 N_f^2\nn
&&\qquad\qquad+
\8{495\02}\left(1+\8{1\06} N_f\right)\left(1-\8{2\033}N_f\right)
\ln{\muMS\02\pi T}\biggr]\left({\alpha_s\0\pi}\right)^{5/2}
+\mathcal O(\alpha_s^3\ln\alpha_s) \biggr\},\quad
\eea
where $\alpha_s=g^2/(4\pi)$ and $\muMS$ is the renormalization scale parameter of
the $\overline{{\rm MS}}$ scheme.
The higher-order terms involve softer scales, but the problem is not
that the effective coupling of these modes is larger. Even
when the running coupling is chosen as that appropriate for hard
modes, the next term, which is of order $g^3$, comes with a
coefficient such that only at temperatures much larger
than $10^5 T_c$ there is apparent convergence of the perturbative series.
At the same time, the dependence on the renormalization scale increases
instead of decreasing,
and this does not get better when higher terms in the perturbative series
are added in, quite to the contrary.

However, as I have already mentioned, this complete failure of the
perturbative expansion is not related to specifically nonabelian
physics. Rather, it can be found in almost any thermal perturbation theory,
such as in the rather trivial large-$N$ limit of scalar field theories.
The latter can be ``solved'' exactly, yielding a rather boring
thermodynamic potential as a function of the coupling, and yet the
perturbative approximations oscillate wildly and seem incapable
of describing the full result except at tiny coupling \cite{Drummond:1997cw}. 
In this example clearly no new physics needs to be incorporated, 
but thermal
perturbation theories should be reorganized. One proposal for
doing so is called ``screened perturbation theory'' 
\cite{Karsch:1997gj,Andersen:2000yj}, 
and has
been generalized to QCD by Andersen, Braaten, and Strickland
\cite{Andersen:1999fw,Andersen:2002ey}.
An alternative proposal, by Blaizot, Iancu, and myself, is
based on an expression for the entropy density that can be
obtained from a $\Phi$-derivable two-loop approximation
\cite{Blaizot:1999ip,Blaizot:2000fc}
(see also \cite{Peshier:2000hx}).
These approaches indeed succeed in taming the plasmon term $\sim g^3$
that spoils the apparent convergence of strict perturbative expansions
in $g$. Moreover, in Refs.~\cite{Blaizot:1999ip,Blaizot:2000fc}
the lattice results for the thermodynamic potential for
$T \ge 3 T_c$ were quite well reproduced by transforming the
leading-order interactions into
so-called hard-thermal-loop \cite{Braaten:1990mz} quasiparticle
properties (which include Debye screening). 

The potential for improvements of the perturbative results
through partial resummations can also be seen in the
a priori strictly perturbative framework of dimensional reduction. 
Using these techniques, the perturbative
expansion of the QCD thermodynamic potential has by now
been carried out up to and including order $g^6 \ln(g)$.
The result is \cite{Kajantie:2002wa}
\bea\label{Pg6}
P\Big|_{g^6\ln g}&=&
\8{8\pi^2\045}T^4 \biggl[1134.8+65.89 N_f+7.653 N_f^2\nn&&
-\8{1485\02}\left(1+\8{1\06} N_f\right)\left(1-\8{2\033}N_f\right)
\ln{\muMS\02\pi T}\biggr]\left({\alpha_s\0\pi}\right)^{3}
\ln{1\0\alpha_s}\,.
\eea

Including this term in a 
strictly perturbative expansion only worsens the
already disastrous apparent convergence.
However, the results (\ref{Fpt}) and (\ref{Pg6}) each involve
contributions from quite different sources. One contribution
is from hard modes, and this behaves decently as concerns
apparent convergence and renormalization scheme dependence.
The problematic part is from the soft sector, which is governed
by an effective three-dimensional Yang-Mills theory with the
chromoelectric degrees of freedom turned into
adjoint scalars with mass $m_E$. Its contribution to four-loop
order is
\bea\label{P3s}
&&P_{\rm soft}/T = {2\03\pi}m_E^3-{3\08\pi^2}\left(
4\ln{\muMS\02m_E}+3\right)g_E^2 m_E^2\nn&&
-{9\08\pi^3}\left({89\024}-{11\06}\ln2+{1\06}\pi^2\right)
g_E^4\,m_E^{\phantom4}
\nn&&
+ N_g {(N g_E^2)^3 \0(4\pi)^4}
\left[ \left( \8{43\012}-\8{157\pi^2\0768} \right) \ln{\muMS\0g_E^2}
+ \left( \8{43\04}-\8{491\pi^2\0768} \right) \ln{\muMS\0m_E}+c \right],
\eea
where $c$ is a constant that is inherently nonperturbative and
needs 3-d lattice calculation to be determined.

\begin{figure}[htb]
\begin{center}\leavevmode
\epsfysize=3cm
\epsfbox{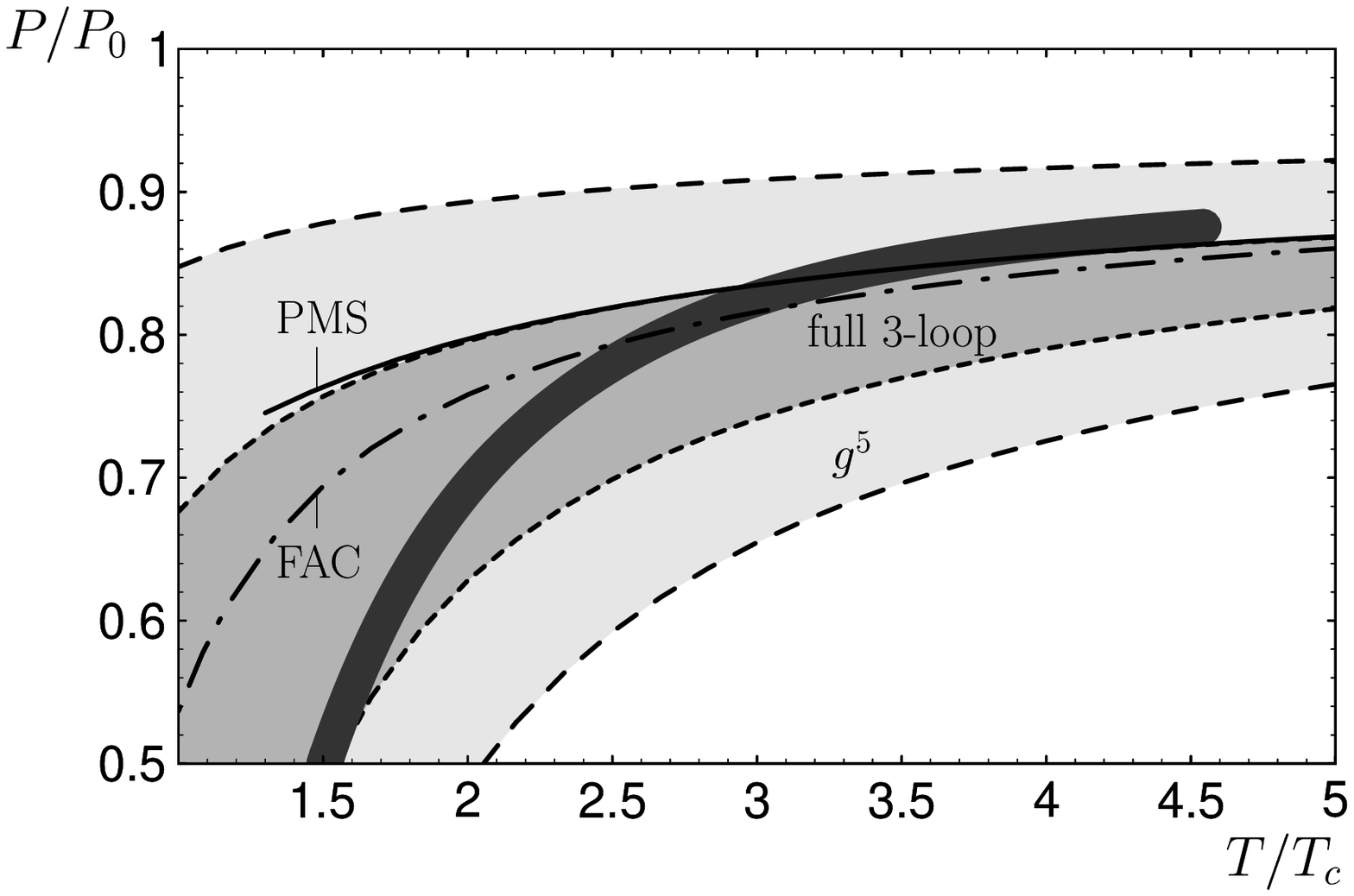}
\qquad
\epsfysize=3.1cm
\epsfbox{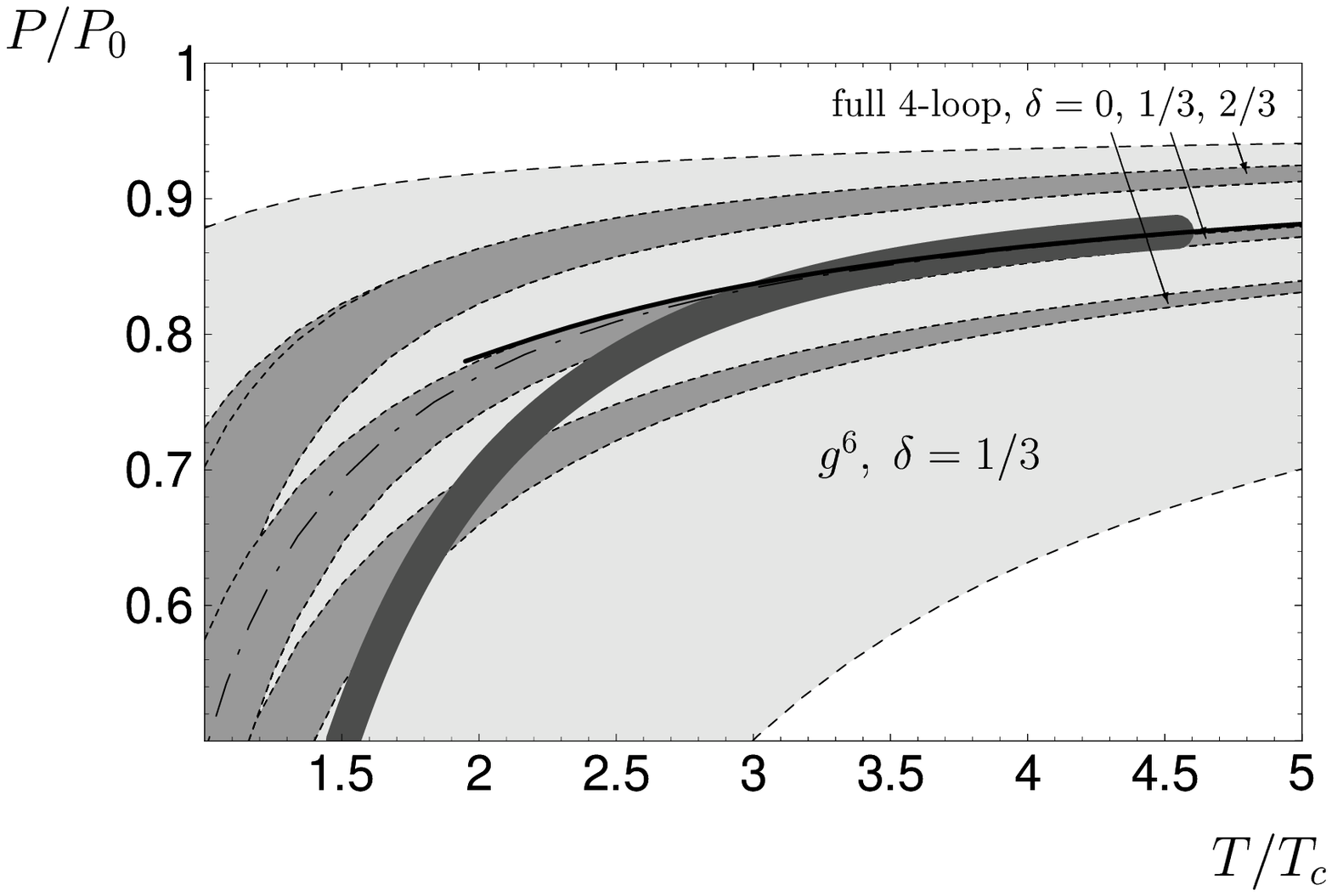}
\end{center}
\vspace*{-0.5cm}
\centerline{\footnotesize\it (a) \hspace{5cm} (b)}
\vspace*{-0.4cm}
\caption[]{{\it (a)} Three-loop pressure in pure-glue QCD with 
unexpanded effective-field-theory parameters 
when $\muMS$ is varied between $\pi T$ and $4\pi T$ (medium-gray band).
The broad light-gray band underneath is the strictly perturbative result to
order $g^5$ with the same scale variations. The full line gives the
result upon extremalization (PMS) with respect to $\muMS$ (which does
not have solutions below $\sim 1.3T_c$); the dash-dotted line corresponds
to fastest apparent convergence (FAC) in $m_E^2$.
{\it (b)} 
The same extended to four-loop order by
including the recently determined $g^6\ln(1/g)$ contribution of
\cite{Kajantie:2002wa} together with three values for
the undetermined constant $\delta$ in $[g^6\ln(1/g)+\delta]$.
The broad light-gray band underneath is the strictly perturbative result to
order $g^6$ corresponding to the central value $\delta=1/3$,
which has a larger scale dependence than the order $g^5$ result; 
the untruncated results on the other
hand show rather small scale dependence.
The full line gives the untruncated
result with $\delta=1/3$ and $\muMS$ fixed by PMS (which does
not have solutions below $\sim 1.9T_c$); the dash-dotted line corresponds
to fastest apparent convergence (FAC) in $m_E^2$.
(Taken from \cite{Blaizot:2003iq})
}
\label{fig34}
\end{figure}

The effective field theory parameters $m_E$ and $g_E$ can
be calculated by perturbative matching, and if their dependence
on $g$ is expanded out and truncated at a given perturbative
order, the bad convergence properties discussed above arise.

\begin{figure}[h]
\begin{center}\leavevmode
\epsfysize=5cm
\epsfbox{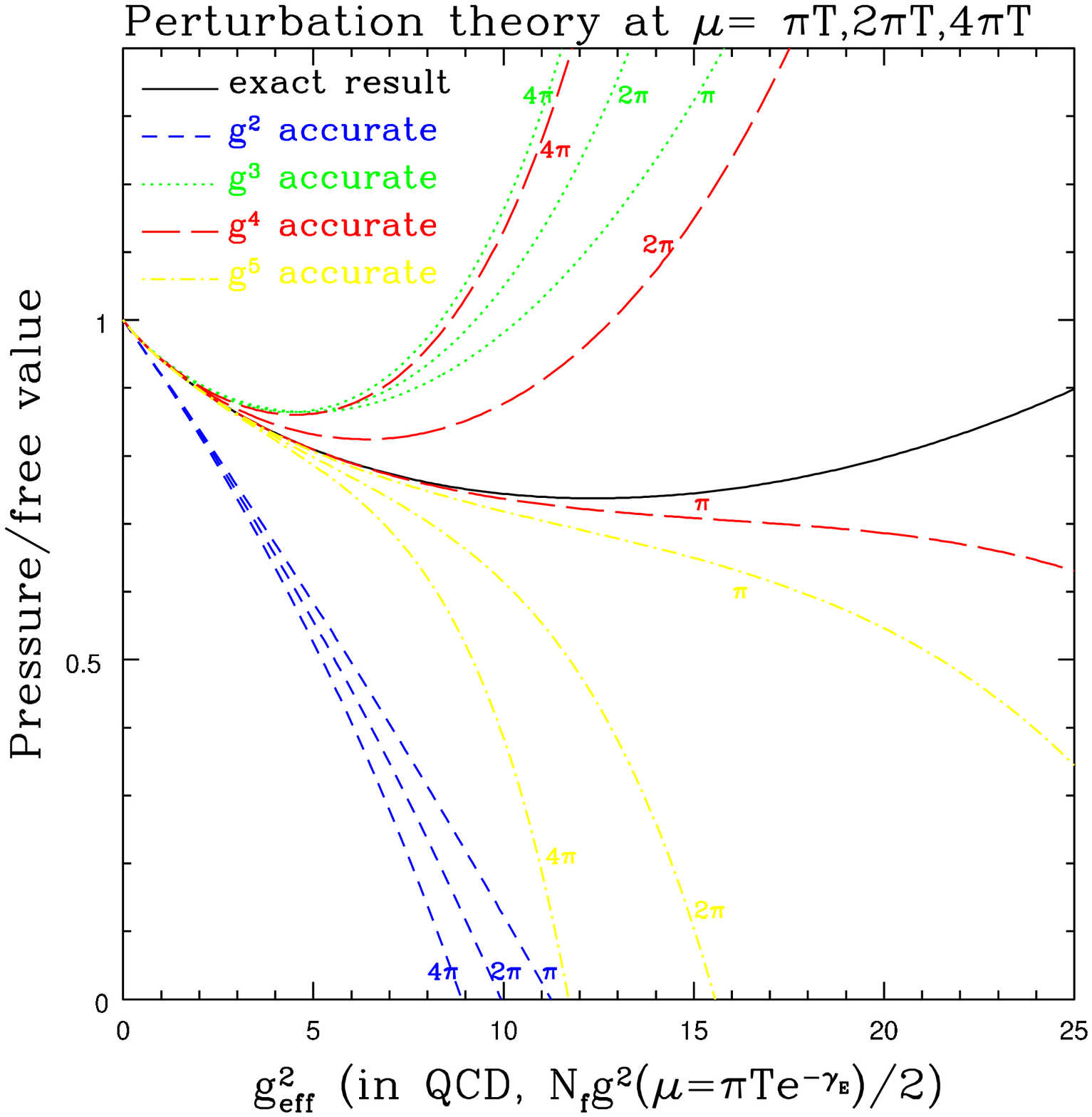}
\end{center}
\vspace*{-0.4cm}
\centerline{\footnotesize\it (a)}
\vspace*{-0.2cm}
\begin{center}\leavevmode
\epsfysize=4.5cm
\epsfbox{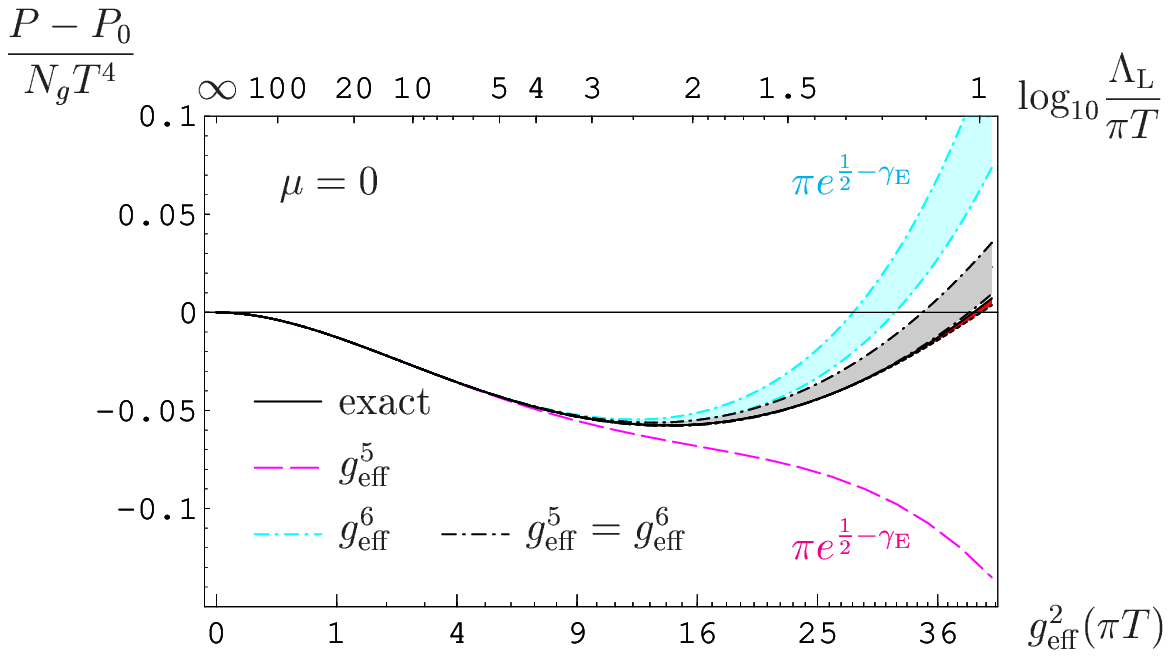}
\end{center}
\vspace*{-0.4cm}
\centerline{\footnotesize\it (b)}
\vspace*{-0.2cm}
\caption[]{{\it (a)} Comparison of the exact result for the pressure
of large-$N_f$ QCD with perturbation theory 
to order $g^2$, $g^3$, $g^4$, and $g^5$, respectively, at
different renormalization points (taken from \cite{Moore:2002md});
{\it (b)} same exact result,
but as a function of $\g^2(\bar\mu_{\rm MS}=\pi T)$ and, alternatively,
$\log_{10}(\Lambda_{\rm L}/\pi T)$. The purple dashed line is
the perturbative result when the latter is evaluated
with renormalization scale
$\bar\mu_{\rm MS}=\bar\mu_{\rm FAC}\equiv\pi e^{1/2-\gamma}T$;
the blue dash-dotted lines include the numerically determined
coefficient to order $\g^6$ (with its estimated error)
at the same renormalization scale.
The result marked ``$\g^5=\g^6$'' 
corresponds to choosing $\bar\mu_{\rm MS}$ such
that the order-$\g^6$ coefficient vanishes and retaining
all higher-order terms contained in the plasmon term $\propto m_E^3$.
The (tiny) red band appearing
around the exact result at large coupling displays the
effect of varying the cut-off from 50\% to 70\% of the
Landau-pole scale $\Lambda_{\rm L}$ (taken from \cite{Ipp:2003jy}).}
\label{figlnf}
\end{figure}

However, the simple prescription of keeping effective field theory parameters
unexpanded in $g$ leads to considerable 
improvements \cite{Blaizot:2003iq}, which are
most striking when the term of order $g^6 \ln(g)$ is included
and the undetermined constant under this log in
(\ref{Pg6}) is fixed. This can be
seen in
Fig.~\ref{fig34}, which even suggests a definite prediction for this
constant under the log, which hopefully will one day be determined
by a combination of analytical and lattice techniques.

A test bed for resummations of perturbative results
which has many of the special features of gauge theories
at finite temperature and/or chemical potential is the
large flavour-number limit of QED or QCD, where the
thermal pressure can be calculated exactly
\cite{Moore:2002md,Ipp:2003zr,Ipp:2003jy}.
Fig.~\ref{figlnf}a shows the lack of convergence of a strict
expansion of the thermodynamic pressure in powers of $g$ and
the large renormalization-point dependences.
Fig.~\ref{figlnf}b on the other hand shows that the above-mentioned
prescription of keeping unexpanded the effective field theory parameters in the
dimensional reduction result to order $g^6$ reproduces the exact result
surprisingly well for all couplings for which the large-$N_f$ theory
is still well-defined (at the highest coupling shown the Landau
pole of this theory is being approached).

\section{Finite quark chemical potential}

\begin{figure}[b]
\begin{center}\leavevmode
\epsfysize=4.1cm
\epsfbox{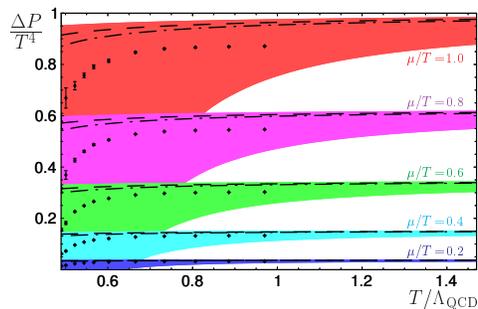}
\end{center}
\vspace*{-0.7cm}
\caption[]{The difference $\Delta P=P(T,\mu)-P(T,0)$ divided by $T^4$ using
the unexpanded three-loop result from dimensional reduction
of Ref.~\cite{Vuorinen:2003fs} for $\mu/T=0.2,\ldots,1.0$
(bottom to top).
Shaded areas correspond to a variation
of $\muMS$ around the FAC-m choice by a factor of 2; dashed and
dash-dotted lines
correspond to the FAC-g and FAC-m results, respectively. 
Also included are the recent lattice data of Ref.~\cite{Allton:2003vx}
(not yet continuum-extrapolated!) assuming $T_c^{\mu=0}=
0.49\, \Lambda_{\rm QCD}$.}
\label{figirv}
\end{figure}

In recent years there has also been substantial progress in
exploring the equation of state of QCD at high temperature and
nonvanishing quark chemical potential on the lattice
\cite{Fodor:2002km,Allton:2003vx,Laermann:2003cv}. These results do not yet
have the quality of results at zero chemical potential, and no
reliable continuum extrapolations have been carried out yet.
But it is plausible that these results are already approximating
continuum results with errors of less than about 20\%. Comparing
HTL-quasiparticle resummation results \cite{Peshier:1999ww,Rebhan:2003wn}
or improved results from dimensional reduction techniques 
extended to finite chemical potential 
\cite{Hart:2000ha,Vuorinen:2002ue,Vuorinen:2003fs} with these
lattice data is in fact most encouraging. Even at the rather
low temperatures that are involved, the analytical weak-coupling
results reproduce the equation of state within expected errors,
see Fig.~\ref{figirv} \cite{Ipp:2003yz}.

There has been for some time a discrepancy between lattice results
on off-diagonal quark-number susceptibilities (i.e.\ derivatives
with respect to chemical potentials for different quark flavours)
and the perturbative result, whose leading-order term 
turns out to be of order $g^6 \ln(g)$ \cite{Blaizot:2001vr}. The authors of
Refs.~\cite{Gavai:2001ie,Gavai:2002kq} have obtained
results in the deconfined phase
that were far below those predicted by perturbation theory,
and have interpreted this as new evidence for nonperturbative
physics and the failure of weak-coupling methods. 
Most recent lattice results \cite{Bernard:2004je,Gavai:2004sd,Allton:2005gk}
have disproved the previous ones, and there is now
agreement with the perturbative estimate at $T\ge 2 T_c$.

With lattice methods, it is possible to explore only
moderate quark chemical potentials at high temperature.
Using weak-coupling techniques and extrapolations down from asymptotically
large chemical potential, much progress has been made in recent years
in unravelling the rich physics of QCD at low temperatures and high
quark chemical potential, which seems to involve numerous different phases
of colour superconductivity \cite{Rajagopal:2000wf,Alford:2001dt,Rischke:2003mt}.

\begin{figure}[tb]
\begin{center}\leavevmode
\epsfysize=4.1cm
\epsfbox{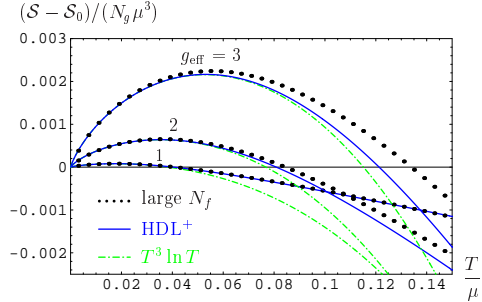}
\end{center}
\vspace*{-0.7cm}
\caption[]{Complete entropy density
in the large-$N_f$ limit for the three values
$\g(\bar\mu_{\rm MS}=2\mu)=1,2,3$ (heavy dots), compared with
the full HDL result (solid line) and the
low-temperature series up to and including the 
$T^3 \ln T$ contributions \cite{Gerhold:2004tb}.}
\label{figentropy}
\end{figure}

Above the critical temperature of colour superconductivity, which
is expected to be somewhere between 6 and 60 MeV
\cite{Rischke:2003mt}, the thermodynamic
behaviour is in fact still fundamentally different from that at high temperatures.
The only weakly screened chromomagnetic interactions lead to
non-Fermi-liquid behaviour \cite{Son:1998uk,Brown:2000eh,Boyanovsky:2000bc,Wang:2001aq,Schafer:2004zf}
which among other effects leads to
an anomalous specific heat when $T$ becomes comparable to
or smaller than $g\mu$. Whereas this effect has been
discussed in QED a long time ago \cite{Holstein:1973} with the conclusion
that this effect is probably unobservably small, in QCD the corresponding
effect is orders of magnitudes larger, since the strong structure
constant times the number of gluons is much larger than the fine structure
constant times one. Until recently, however, the corresponding
resummed perturbation series has been unavailable except for the
coefficient in front of a
leading logarithmic term \cite{Chakravarty:1995}. 
This has been completed in \cite{Ipp:2003cj,Gerhold:2004tb}, 
leading to a low-temperature 
entropy with a perturbative expansion involving logs and fractional
powers of the temperature of the form
\bea\label{anomS}
&&\mathcal S/\mu^2 T = N_f
+{4\alpha_s N_f\09\pi} 
\ln\left(2.227
\sqrt{\alpha_s N_f\0\pi}{\mu\0T}\right) 
-1.383\left({\alpha_s N_f\0\pi}\right)^{2\03} 
\left(T\0\mu\right)^{2\03}\nonumber\\
&&\qquad \qquad 
+1.041\left({\alpha_s N_f\0\pi}\right)^{1\03} 
\left(T\0\mu\right)^{4\03}
+ O(T^3 \ln T),\quad T\ll g\mu.
\eea

Using hard-dense-loop (HDL) resummations 
\cite{Altherr:1992mf,Vija:1995is,Manuel:1996td},
quantitative results are also available for $T\sim g\mu$ 
\cite{Gerhold:2004tb}
and these can be tested against
the solvable case of large-$N_f$ QCD, as shown in Fig.~\ref{figentropy}.
Physically interesting applications of these results are presumably
outside the reach of heavy-ion colliders. But they may be of
interest to neutron star physics, namely the cooling behaviour
of young neutron stars if they have a normal-conducting quark-matter component.
Both the specific heat and the neutrino emissitivity contain large
logarithms from non-Fermi-liquid behaviour, with important deviations
\cite{Schafer:2004jp,Gerhold:2005uu}
from naive lowest-order perturbation theory \cite{Iwamoto:1980eb}.

\section{Conclusions}\label{concl}

To summarize, there has been quite some recent progress in
evaluating by weak-coupling techniques the thermodynamics
of a deconfined quark-gluon plasma. At zero chemical potential,
the calculation has been pushed to order $g^6 \log(g)$, and
at large chemical potential analytical results exist for
the non-Fermi-liquid behaviour of ungapped quark matter and
also of colour superconductors. At zero to small chemical
potential, comparisons to lattice results are possible and
for $T\ge 2 T_c$
seem to validate the outcome of resummations that are necessary
to overcome the poor apparent convergence of strict perturbation
theory.

 \section*{Acknowledgments}
I would like to thank the organizers of the 4th Budapest Winter School 
On Heavy Ion Physics for their warm hospitality. Part of the
work presented here has been supported by the Austrian Science
Foundation FWF, project no.\ P16387.
 


\end{document}